\documentclass[12pt]{article}
\pdfoutput=1
\usepackage[utf8]{inputenc}
\usepackage[english]{babel}
\usepackage{amssymb,graphicx}
\usepackage{amsmath,amsfonts}
\usepackage{fullpage}

\newcommand{\be}{\begin{equation}}
\newcommand{\ee}{\end{equation}}
\newcommand{\bea}{\begin{eqnarray}}
\newcommand{\eea}{\end{eqnarray}}

\title{\vspace{-2cm} 
\begin{flushright}
{\normalsize INR-TH-2019-013}
\end{flushright}
\vspace{0.5cm} 
New constraints on Lorentz Invariance violation from Crab Nebula spectrum beyond $100$ TeV}
\author{Petr Satunin\thanks{{\bf e-mail}: satunin@ms2.inr.ac.ru}
\vspace{.2cm}\\
\normalsize\it  Institute for Nuclear Research of the Russian Academy of Sciences, \\  
\normalsize \it  60th October Anniversary Prospect, 7a, 117312  Moscow, Russia} 
\date{}
\begin{document}
\maketitle

\begin{abstract}
Recently two collaborations, Tibet and HAWC, presented new measurements of gamma-ray spectrum from Crab Nebula \cite{Amenomori:2019rjd},\cite{Abeysekara:2019edl} which continues beyond $100$ TeV. We use these data to establish two-sided constraints on parameters of Lorentz Invariance violation in quantum electrodynamics.  The limits on Lorentz violating mass scale for quartic dispersion relation are $4.1\times 10^{14}\ \mbox{GeV}$ (photon splitting) and $1.9\times 10^{13}\ \mbox{GeV}$ (photon decay) for superluminal case, and $1.4\times 10^{12}$ GeV (suppression of shower formation) for subluminal case. 
\end{abstract}

The Crab Nebula, pulsar wind nebula which is a remnant from supernova SN 1054, is one of the brightest and most studied galactic gamma ray sources. Since 1989 when the first TeV gamma rays from Crab Nebula were detected by Whipple collaboration \cite{Weeks}, the Crab Nebula remains the source with the most energetic detected photons.  In 2004, HEGRA collaboration reported the Crab Nebula spectrum collected over more than 10 years of operation  \cite{Aharonian:2004gb}. In that report, the detection of $75$ TeV photons was established with statistical significance $2.7$ sigma. It took 15 years to extend the measurements to higher energies. This year two collaborations, Tibet and HAWC, presented the highest-energy Crab Nebula spectra \cite{Amenomori:2019rjd},\cite{Abeysekara:2019edl}; both spectra continue beyond 100 TeV. Besides improving the knowledge about the source, this detection allows us to better constrain  some scenarios of new physics such as hypothetical violation of Lorentz Invariance (LI).


Violation of LI (LV for short) is motivated by several approaches to gravity quantization (see reviews \cite{Mattingly:2005re,Liberati:2013xla}  and references therein) and usually considered in the matter sector in the framework of effective field theory \cite{Coleman:1997xq,Colladay:1998fq,Jacobson:2002hd,Myers:2003fd,Kostelecky:2009zp}.
LV in the photon sector modifies several processes responsible for creation, propagation and detection of photons.
These include photon decay \cite{Coleman:1997xq, Jacobson:2002hd,Rubtsov:2012kb}, photon splitting \cite{Gelmini:2005gy, Astapov:2019xmt} and suppression of the Bethe-Heitler process \cite{Vankov:2002gt, Rubtsov:2012kb, Rubtsov:2016bea}. Most of these effects would lead to a significant reduction of the observed photon flux, which is not seen in the data. However, some of the effects may lead to an increase in the observed photon flux, such as the suppression of pair production on background lights \cite{Kifune:1999ex, Stecker:2001vb}.

We specify ourselves to the following model\footnote{We assume that gauge, rotational and CPT symmetries are unbroken and consider LV operators of dimension larger than $4$; additional requirements to the model are gathered in \cite{Rubtsov:2012kb}. LV in the electron sector is not considered here since those constraints are more stronger than in the photon sector \cite{Liberati:2012jf}, see also discussion in \cite{Rubtsov:2016bea}.},
\begin{equation}
\label{L1}
\mathcal{L}=-\frac{1}{4}F_{\mu\nu}F^{\mu\nu}  \mp \frac{1}{2 M_{LV}^2}F_{ij}\Delta^2 F^{ij}  +i\bar{\psi}\gamma^\mu D_\mu\psi - m\bar{\psi}\psi.
\end{equation}
In comparison with the standard QED Lagrangian, Eq. (\ref{L1}) contains a single extra LV term, suppressed by a certain mass scale $M_{LV}$\footnote{The mass scale $M_{LV}$ corresponds to the parameter $c^{(6)}_{(I)00}$ of the most general model called non-minimal Standard Model Extension (SME)  
\cite{Kostelecky:2009zp}, $c^{(6)}_{(I)00} = - \sqrt{\pi}/M_{LV}^{2}$.}, which is usually considered to be connected with the scale of quantum gravity.  The Lagrangian (\ref{L1}) leads to a modification of the photon dispersion relation,
\begin{equation}
\label{DispRel}
E_\gamma^2 = k_\gamma^2 \pm \frac{k_\gamma^4}{M_{LV}^2}.
\end{equation}
The sign "$+$" in the dispersion relation is connected with superluminal case, while the sign "$-$" --- with subluminal\footnote{The subluminal type of LV for photons (eq.(\ref{DispRel}), sign "$-$") may be induced by radiative corrections caused by any charged particle with nonzero LV operators of dimension $4$ \cite{Satunin:2017wmk}.}
. 
The most important processes for superluminal case are photon decay $\gamma \to e^+e^-$ and photon splitting $\gamma \to 3\gamma$.
Thus, a photon propagating from Crab Nebula to Earth, may decay via these  two channels so the photon flux from Crab reduces before reaching Earth. 
On the other hand, in the subluminal case a photon lacks energy which suppresses the pair production on nuclei (Bethe-Heitler process), allowed in the LI case. This process is crucial for the formation of atmosphere showers used to detect TeV gamma-rays. Its suppression will again lead to a reduction of the measured flux.

\paragraph{Photon decay} The photon decay $\gamma \to e^+e^-$ is a threshold process, which switches on if the effective photon mass $m_{\gamma, eff} \equiv \sqrt{E_\gamma^2-k_\gamma^2} = E_\gamma^2/M_{LV}$ is larger than twice the electron mass, $m_{\gamma, eff} > 2m_e$. Once being allowed, the decay is very fast \cite{Rubtsov:2012kb} so no photons with energy above the threshold  reach Earth.  Thus, even a single photon event with energy $E_\gamma$ constrains $M_{LV}$ to lie above
\begin{equation}\label{PhotDecay}
M_{LV} \geq \frac{E_\gamma^2}{2m_e}\;.
\end{equation}
The statistical significance of the constraint coincides with the significance of the corresponding photon event.  The current constraint on $M_{LV}$ from the absence of photon decay is $M_{LV}>2.8\times 10^{12}$ GeV \cite{Martinez-Huerta:2016azo}. 

\paragraph{Photon splitting}
Another channel of the photon decay is the triple photon splitting $\gamma \to 3\gamma$. This process does not have a threshold and occurs whenever LV is superluminal. Due to the phase volume suppression, the width is small but nonzero \cite{Astapov:2019xmt},
\begin{align}
\label{FinalGamma}
\Gamma_{\gamma\to 3\gamma}\ &\simeq\ \,1.2\cdot 10^3\left(\frac{2\alpha^2}{45}\right)^2\frac{1}{2^8\,3!\,\pi^4}\frac{E_{\gamma}^{19}}{m_e^8M^{10}_{LV}} \simeq \notag \\ &\simeq\ 5\cdot 10^{-14}\;\frac{E_{\gamma}^{19}}{m_e^8M^{10}_{LV}}.
\end{align} 
Note the strong dependence of the width on energy. 

The probability for a photon not to split while traveling from Crab to Earth obeys exponential distribution,
\begin{equation}
\label{P}
P= \mbox{e}^{-\Gamma_{\gamma\to 3\gamma}\;\times \; L_{CRAB}},
\end{equation}
where $L_{CRAB}=2$ kpc is the distance from Crab to Earth. The factor $P$ denotes the suppression of a photon flux compared to the standard LI case,
\begin{equation}
\label{LVhyp}
\bigg(\frac{d\Phi}{dE}\bigg)_{LV}= P \times \bigg(\frac{d\Phi}{dE}\bigg)_{LI}\;.
\end{equation}
The predicted photon flux $\left(\frac{d\Phi}{dE}\right)_{LV}$ can be tested against experimental data points. As a result of the test, a certain constraint on the suppression factor $P$ will be established. The latter, following eqs.(\ref{FinalGamma}),(\ref{P}), transfers to the constraint on the mass scale $M_{LV}$,  
\begin{equation}
\label{SplippingGen}
M_{LV} > 2.3\times 10^{14}\;\mbox{GeV}\cdot \left( \frac{E}{100\,\mbox{TeV}}\right)^{1.9} \cdot \left( \frac{1}{-\log P}\right)^{0.1}, 
\end{equation}
for the fixed value of $E$. Thus, the last bin of HEGRA data \cite{Aharonian:2004gb}, centered at $E=75$ TeV, gives the constraint $M_{LV}>1.3\times 10^{14}$ GeV \cite{Astapov:2019xmt}. We are going to see that the new data above $100$ TeV photon energy allow us to improve this constraint.

\paragraph{Shower formation}
Subluminal type of LV predicts the suppression of Bethe-Heitler process \cite{Rubtsov:2012kb} which is responsible for the first interaction of an astrophysical photon in the atmosphere. Thus, in this case atmospheric showers initiated by photons would be deeper than in the standard case \cite{Rubtsov:2016bea}. Very deep showers would escape registration in the experiment. Thus, the prediction for subluminal LV is similar to superluminal case: the suppression of photon flux for highest-energy photons.

If the depth $X_0$ of the photon first interaction in the atmosphere is larger than the total atmosphere depth $X_{\rm atm}$\footnote{Calculated taking into account the altitude and the maximal zenith angle of detection of the experiment.}, the shower will not develop, and the event will not be detected. The probability for a photon to produce pair in the atmosphere reads, 
\begin{equation}
\label{Preg}
P=\int_0^{X_{\rm atm}}dX_0
\;\frac{{\rm e}^{-X_0/\langle X_0\rangle_{LV}}}{\langle X_0\rangle_{LV}}\; = 1 - {\rm e}^{-X_{\rm atm}/\langle X_0\rangle_{LV}},
\end{equation} 
where the mean depth of the first interaction for LV case $\langle X_0\rangle_{LV}$ is expressed via LI mean depth $\langle X_0\rangle_{LI}=57$ g $\mbox{cm}^{-2}$, and the ratio of the Bethe-Heitler cross-sections in the standard and Lorentz violating theories,  
\begin{equation}
\label{X0LV}
\langle X_0\rangle_{LV}=\frac{\sigma_{\rm BH}}{\sigma^{\rm LV}_{\rm BH}}
\langle X_0\rangle_{LI}\;.
\end{equation} 
The latter is calculated in \cite{Rubtsov:2012kb},
\begin{equation}
\label{suppr}
\frac{\sigma^{\rm LV}_{\rm BH}}{\sigma_{\rm BH}}\simeq 
\frac{12 m_e^2 M_{LV}^2}{7E_\gamma^4}\cdot 
\log\frac{E_\gamma^4}{2m_e^2 M_{LV}^2};
\end{equation}
the expression inside the log in (\ref{suppr}) we call $A$.  As for the case of photon splitting, the detected photon flux from Crab Nebula would be suppressed as in (\ref{LVhyp}) with $P$  expressed by (\ref{Preg}). The absence of such suppression in the data yields the constraint on $P$, which, in turn,  transfers to the constraint on $M_{LV}$ in the following way,
\begin{align}
\label{AAA}
&\frac{A}{\log A} < \frac{12.78}{-\log (1-P)}, \\ &A \equiv \frac{E^4}{2m_e^2 M_{LV}^2}= 1.9\cdot \left( \frac{E}{100\,\mbox{TeV}}\right)^4 \cdot \left( \frac{M_{LV}}{10^{13}\,\mbox{GeV}} \right)^{-2}.\notag
\end{align}
The eq.(\ref{AAA}) is solved numerically for fixed $P$.
The bound obtained from HEGRA data \cite{Aharonian:2004gb} reads $M_{LV}>2.1\times 10^{11}$ GeV \cite{Rubtsov:2016bea}. The suppression grows with energy, see (\ref{suppr}),(\ref{AAA}) so we can expect stronger constraints from Tibet and HAWC.







\paragraph{Tibet.} The Tibet collaboration has published the combined data from air shower ground array of detectors and underground array of muon detectors collected during $719$ days of observation \cite{Amenomori:2019rjd}. The altitude of Tibet array is $4300$ m above the sea level, so the depth of the atmosphere at the Tibet location is not more than $780$ g $\mbox{cm}^{-2}$ for showers from the maximal zenith angle $40$ degrees (events with larger zenith angles have not been considered in Tibet analysis \cite{Amenomori:2019rjd}).

The statistical significance for each energy bin of Crab nebula photon spectrum was calculated by the likelihood ratio method following Li \& Ma \cite{Li:1983fv}. The last but one energy bin of Tibet data \cite{Amenomori:2019rjd} (energy range $100$-$250$ TeV, median energy $140$ TeV) contains $N_{on}=20$ on-source and  $N_{off}=94$ off-source photon events\footnote{In the article \cite{Amenomori:2019rjd} the joint number of events in two last bins, and in the last bin are presented; these numbers are just subtraction.}; the ratio of on-source and off-source exposures is $\alpha=0.05$, the number of signal events is $N_s=N_{on}-\alpha N_{off}=15.3$. The calculated statistical significance is $5.0\,\sigma$. The last energy bin ($250$-$630$ TeV) contains only $4$ photon-like on-source events\footnote{Moreover, one of these events may be a cosmic ray event with a probability of $0.23$\cite{Amenomori:2019rjd}.}, the corresponding statistical significance is $2.4\,\sigma$. The statistics in the last bin is too low to infer any significant bounds on LV, so in our analysis we use the last but one bin.
 

We test the hypothesis that the photon flux (i.e. the number of signal events\footnote{The coefficient of proportionality between the number of signal events and the photon flux is determined by their ratio in the last but one energy bin, $N_s=15.3$ and $\langle \frac{d\Phi}{dE}\rangle =2.4\cdot 10^{-17}\ \mbox{TeV}^{-1}\mbox{cm}^{-2}\mbox{s}^{-1}$.}) is suppressed by a factor $P$.
The expectation value for the signal events $\langle N_s \rangle^{LI}$ is obtained by extrapolation of power-law fit of the low energy part of the spectrum (less than $20$ TeV), to high energies. To be conservative, we use the power-law fit of HEGRA \cite{Aharonian:2004gb} data,
\begin{equation}
\label{HegraSp}
\left( \frac{d\Phi}{dE}\right)_{LI}=2.83 \cdot 10^{-11}\cdot(E/\mbox{TeV})^{-2.62}\ \mbox{TeV}^{-1}\mbox{cm}^{-2}\mbox{s}^{-1}.
\end{equation}
 In the presence of LV the expected signal gets suppression $P$, $\langle N_s \rangle^{LV} = P\times \langle N_s \rangle^{LI}$. In order to obtain the probability of the observed realization ($N_{on}, N_{off}$) for the expectation number of the signal events $\langle N_s \rangle^{LV}$ we use likelihood ratio method, marginalizing over unknown background; the details are similar to those presented in  \cite{Rubtsov:2016bea}.
As a result, the suppression factor $P=0.17$ is excluded at $95\%$ CL.  


As we mentioned before, the suppression factor $P$ may be caused either by the photon splitting or Bethe-Heitler suppression. For numerical results we take $E=140$ TeV. The constraint (\ref{SplippingGen}) from the absence of photon splitting (superluminal case) reads,
\begin{align}
\label{TibetSpli}
\mbox{(superluminal)}\quad M_{LV} > 4.1 \times 10^{14}\; \mbox{GeV},  \qquad 95\%\; \mbox{CL}.
\end{align}
The constraint (\ref{AAA}) from non-suppression of the Bethe-Heitler process (subluminal case) reads
,
\begin{equation}
\label{BoundBetheHeitlerTibet}
\mbox{(subluminal)}\qquad M_{LV} > 1.4 \times 10^{12}\; \mbox{GeV}, \qquad 95\%\; \mbox{CL}.
\end{equation}

Let us also give the constraint from the photon decay $\gamma \to e^+e^-$. The bound (\ref{PhotDecay}) applied to the bin of Tibet data centered at $E=140$ TeV, reads (remember that the significance of the bin gives the significance for the bound)
\begin{equation}
\label{Photon_decay_Tibet}
\mbox{(superluminal)}\quad M_{LV} > 1.9 \times 10^{13}\; \mbox{GeV}, \quad  5\sigma.
\end{equation}
This constraint is an order of magnitude weaker than the splitting constraint (\ref{TibetSpli}). However, the constraint (\ref{Photon_decay_Tibet}) is of very high statistical significance. Moreover, the photon decay bound (\ref{Photon_decay_Tibet})  does not rely on any additional assumption such as maximal zenith angle or the shape of Crab Nebula spectrum.

\paragraph{HAWC}
HAWC observatory is an array of water Cerenkov detectors located in Mexico at the altitude 4100 meters. The maximal atmosphere depth corresponding to the maximal zenith angle $45$ degrees (see \cite{Malone:2018zeg}) is $865\;\,\mbox{g}\;\mbox{cm}^{-2}$. The last energy bin in which Crab Nebula was detected by HAWC \cite{Abeysekara:2019edl}, is $100$-$177$ TeV
. The energy reconstruction is performing two independent methods, "ground parameter" (GP) and neural network (NN).  The reconstructed median energy of the last bin is $102$ and $118$ TeV for two methods respectively. 

The photon decay bound (\ref{PhotDecay}) applied to the median energy of the last bin gives,
\begin{equation}
\label{HAWC_gdecay}
\mbox{(superluminal)}\qquad M_{LV} > 1.0\; (1.4) \times 10^{13}\ \mbox{GeV}.
\end{equation}
Here the first value corresponds to GP method while the value in the brackets --- to NN method.

Since the HAWC collaboration does not provide the details of background \cite{Abeysekara:2019edl}, we are not allowed to perform statistical analysis based on the number of on-source and off-source events. Instead of that we perform analysis based on the photon flux. Assuming Gaussian distribution (which is not in fact true for small number of events) with given mean value and dispersion for the measured flux in the energy bin $100$-$177$ TeV, we 
apply Pearson's chi-squared criterium to test a hypothesis of photon flux $ \left( \frac{d\Phi}{dE}\right)_{LV}=P\cdot \left( \frac{d\Phi}{dE}\right)_{LI}$ against measured flux.
We take  $\left( \frac{d\Phi}{dE}\right)_{LI}$ 
 as power-law extrapolation of HEGRA spectrum (\ref{HegraSp}), the same as for the Tibet data analysis.
As a result, suppression factors excluded at $95\%$ CL read $P=0.09$ and $P=0.18$ for GP and NN method respectively. 

Let us show our estimation for $95\%$ CL bound on $M_{LV}$. First, we start from the splitting constraint (\ref{SplippingGen}) which is connected with superluminal LV. The constraint reads, 
\begin{align}
\label{HAWCsplit}
\mbox{(superluminal)}\ \  M_{LV} > 2.2\; (3.0) \times 10^{14}\ \mbox{GeV}, \ \  95\%\; \mbox{CL},
\end{align}
here two values correspond with two reconstruction methods as previously.
Further, let us provide the estimated constraint (\ref{AAA}) based on the absence of shower suppression (subluminal type of LV)\footnote{Let us also provide the bounds based on chi-squared criterium applied to two last bins of HAWC data instead of the last one: $M_{LV} > 4.4\; (6.9) \times 10^{11}\ \mbox{GeV}, \quad 95\%\; \mbox{CL}$.},
\begin{align}
\label{HAWCshower}
\mbox{(subluminal)}\quad M_{LV} > 4.7\; (9.7) \times 10^{11}\ \mbox{GeV}, \quad 95\%\; \mbox{CL},
\end{align}
The bounds (\ref{HAWCsplit}),(\ref{HAWCshower}) are worse than the Tibet ones because the HAWC median energy is less than the Tibet one, and the statistics is less as well.

\paragraph{Discussion} 
By the analysis of the Crab Nebula spectra reported by Tibet and HAWC collaborations, we have obtained bounds on the LV mass scale  in the photon sector.  For the superluminal type of LV, the best of our constraints which are based on Tibet data, read
\begin{align}
\mbox{photon decay} \qquad & M_{LV} > 1.9 \times 10^{13}\; \mbox{GeV} \qquad \ 5\sigma, \notag \\
\mbox{photon splitting} \quad & M_{LV} > 4.1 \times 10^{14}\; \mbox{GeV}  \quad \ 95\%\; \mbox{CL}. \notag
\end{align}
These bounds are several times better than the previous ones $M_{LV}>1.3\times 10^{14}$ GeV \cite{Astapov:2019xmt} (photon splitting), $M_{LV}>2.8\times 10^{12}$ GeV \cite{Martinez-Huerta:2016azo} (photon decay), based on HEGRA data.  
The splitting constraint is the best in the literature for the superluminal case. Alghough the photon decay constraint is an order of magnitude weaker, it is the most robust bound which does not rely on any astrophysical assumption (intrinsic spectrum of the source, zenith angle, etc.).
 
For the subluminal case we improve the bound from shower formation \cite{Rubtsov:2016bea} with Tibet data by an order of magnitude, 
$$
M_{LV} > 1.4 \times 10^{12}\; \mbox{GeV}, \qquad 95\%\; \mbox{CL}.
$$
It is worth  comparing it with another bound that exists in the subluminal case and arises from pair production by extragalactic photons on extragalactic background light (EBL). In the presence of LV of subluminal type, the TeV photons would propagate through the extragalactic medium without significant suppression, which contradicts observational data \cite{Kifune:1999ex, Stecker:2001vb}.  The current limits on $M_{LV}$ from pair production on EBL are $7.8\times 10^{11}\;\mbox{GeV}$ \cite{Abdalla:2019krx} and  $2.4\times 10^{12}\;\mbox{GeV}$ \cite{Lang:2018yog} (both $95\%$ CL), which are of the same order as our shower suppression constraints (\ref{BoundBetheHeitlerTibet}), (\ref{HAWCshower}). 

There is also another bound of this type, based on current non-observation of ultra-high-energy (UHE) photons (energy $\sim\; 10^{19}$ eV) which are awaited to be one of the products of GZK process \cite{Greisen:1966jv} --- pion production of UHE cosmic rays on cosmic microwave background (CMB). These photons, if created, produce pairs on CMB and radio backgrounds; in the presence of LV of subluminal type the process of pair production is suppressed. Current non-detection of such photons sets the bound $M_{LV} > 10^{22}$ GeV \cite{Galaverni:2007tq,Galaverni:2008yj,Maccione:2008iw,Lang:2017wpe}. However, this bound strictly rely on the chemical composition of cosmic rays, which is still not clear \cite{Abbasi:2015xga}, as well as on the spectral shape and source evolution (see discussion in \cite{Lang:2017wpe}).

Let us note that aforementioned constraints referred only to quartic LV corrections to photon dispersion relation (\ref{DispRel}). However, the constraints referred to EBL suppression, as well as to GZK photons, are made also for cubic correction to dispersion relation. The generalization of the splitting and shower formation bounds to cubic LV is not straightforward, the calculation of corresponding cross-sections in the appropriate model (see \cite{Myers:2003fd}) is necessary.


\paragraph{Aknowledgements} The author thanks Dmitry Kirpichnikov, Grigory Rubtsov, Sergey Sibiryakov and Sergey Troitsky for helpful discussions and Jim Linnemann for useful comments. The research was supported by the state assignment number 0031-2014-0066 of the INR RAS

\end{document}